\documentclass[twocolumn,floatfix,aps,eqsecnum,superscriptaddress,prb]{revtex4-2}
\usepackage{graphicx}
\usepackage{amsmath}
\usepackage{amssymb}
\usepackage{bm}
\usepackage{hyperref}

\begin{document}

\title{Magnus effect on a Majorana zero-mode}
\author{G. Lemut}
\affiliation{Instituut-Lorentz, Universiteit Leiden, P.O. Box 9506, 2300 RA Leiden, The Netherlands}
\author{M. J. Pacholski}
\affiliation{Instituut-Lorentz, Universiteit Leiden, P.O. Box 9506, 2300 RA Leiden, The Netherlands}
\affiliation{Max Planck Institute for the Physics of Complex Systems, N\"{o}thnitzer Strasse 38, 01187 Dresden, Germany}
\author{S. Plugge}
\affiliation{Instituut-Lorentz, Universiteit Leiden, P.O. Box 9506, 2300 RA Leiden, The Netherlands}
\author{C. W. J. Beenakker}
\affiliation{Instituut-Lorentz, Universiteit Leiden, P.O. Box 9506, 2300 RA Leiden, The Netherlands}
\author{\.{I}. Adagideli}
\affiliation{Faculty of Engineering and Natural Sciences, Sabanci University, Orhanli-Tuzla, Istanbul, Turkey}
\affiliation{MESA+ Institute for Nanotechnology, University of Twente, 7500 AE Enschede, The Netherlands}
\affiliation{T\"{U}B\.{I}TAK Research Institute for Fundamental Sciences, 41470 Gebze, Turkey}

\date{March 2023}
\begin{abstract}
A supercurrent on the proximitized surface of a topological insulator can cause a delocalization transition of a Majorana fermion bound to a vortex core as a zero-mode. Here we study the dynamics of the deconfinement, as a manifestation of the Magnus effect (the coupling of the superflow to the velocity field in the vortex). The initial acceleration of the Majorana fermion is $\pm 2v_{\rm F}^2 K/\hbar$, perpendicular to the Cooper pair momentum $\bm{K}$, for a $\pm 2\pi$ winding of the superconducting phase around the vortex. The quasiparticle escapes with a constant velocity from the vortex core, which we calculate in a semiclassical approximation and compare with computer simulations.
\end{abstract}
\maketitle

\section{Introduction}
\label{sec_intro}

A topological superconductor can bind a Majorana fermion as a midgap state in the core of a magnetic vortex \cite{Kop91,Vol99,Fu08}. This Majorana zero-mode has been dubbed the ``Zen particle'' \cite{Frolov}, because it embodies nothingness: it has zero charge, zero spin, zero energy, and zero mass \cite{Bee13,Das15,Lut18}. It does have a definite chirality, set by the sign of the $\pm 2\pi$ winding of the superconducting phase around the vortex \cite{Jac81}. 

A superflow couples to the circulating phase, producing a sideways force on the vortex known as the Magnus force \cite{Noz66,Mak95,Sto96,Son97}. It was recently shown \cite{Pac21} that the superflow also acts on the zero-mode, causing a deconfinement transition when the Cooper pair momentum $K$ exceeds the critical value $\Delta_0/v_{\rm F}$ (with $\Delta_0$ the superconducting gap and $v_{\rm F}$ the Fermi velocity).

Here we follow up on that work and investigate the dynamics of the transition, when the superconductor is quenched by the sudden application of a superflow.  Computer simulations show that the Majorana zero-mode escapes from the vortex core as a wave packet with a constant velocity $v_{\rm escape}$. A key result of our analysis is a calculation of the dependence of this quantity on $K,\Delta_0$, and $v_{\rm F}$, in a semiclassical approximation that is found to agree well with the simulations. That calculation is presented in Sec.\ \ref{sec_semiclassics}, after we have formulated the problem (Sec.\ \ref{sec_quench}) and solved for the short-time dynamics (Sec.\ \ref{sec_short}). We compare with computer simulations in Sec.\ \ref{sec_simulations} and conclude in Sec.\ \ref{sec_conclude}.

\section{Quenched topological superconductor}
\label{sec_quench}

\begin{figure}[tb]
\centerline{\includegraphics[width=0.9\linewidth]{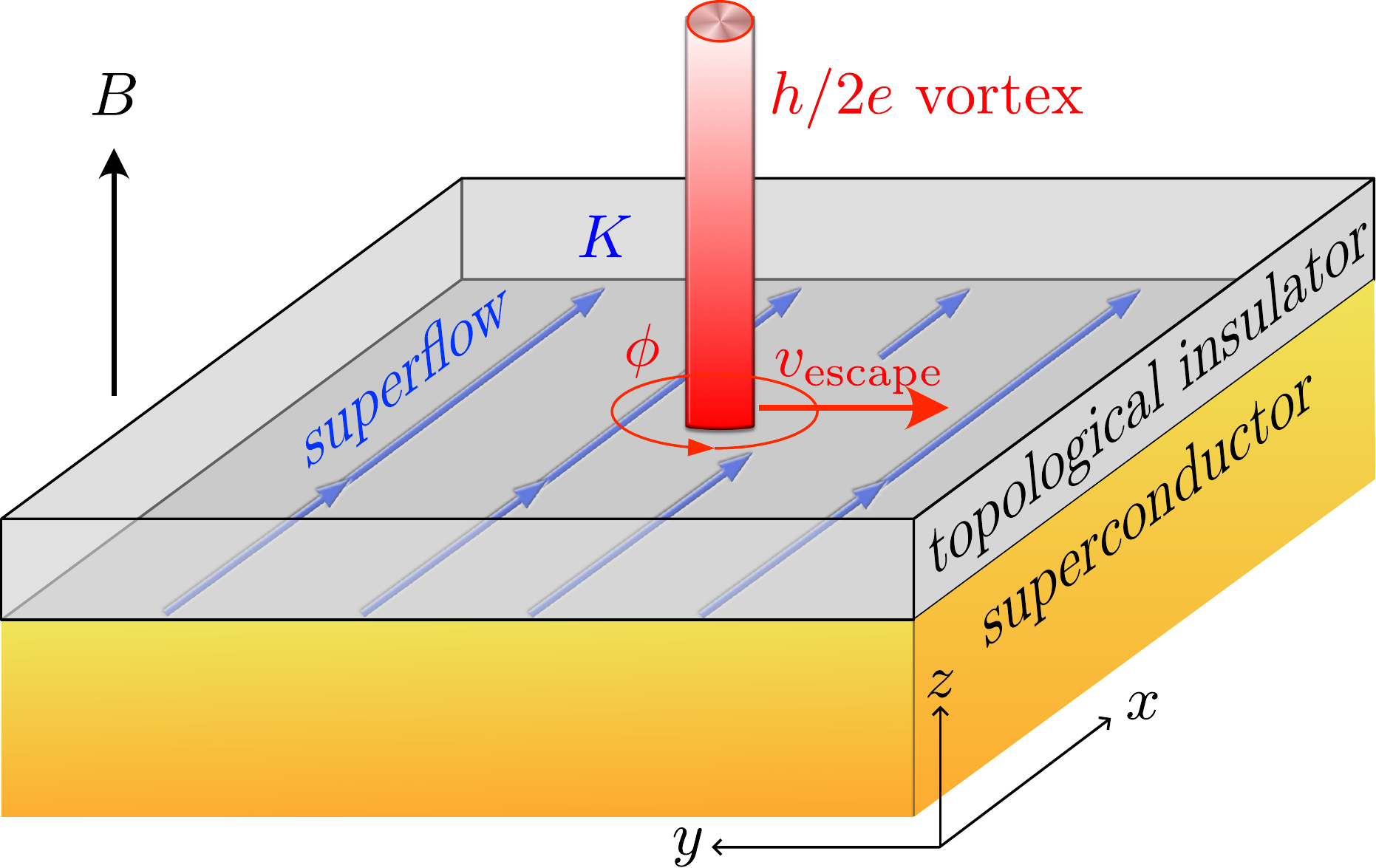}}
\caption{Schematic of a topological insulator with induced superconductivity (gap $\Delta_0$) in a perpendicular magnetic field $B$. A vortex (red, with a $ 2\pi$ winding of the superconducting phase $\phi$) binds a Majorana zero-mode. An in-plane supercurrent (blue arrows, Cooper pair momentum ${K}$) can deconfine the zero-mode, producing a Majorana fermion wave packet that escapes with velocity $ v_{\rm escape}$ in a direction perpendicular to the superflow.
}
\label{fig_layout}
\end{figure}

\begin{figure}[tb]
\centerline{\includegraphics[width=1\linewidth]{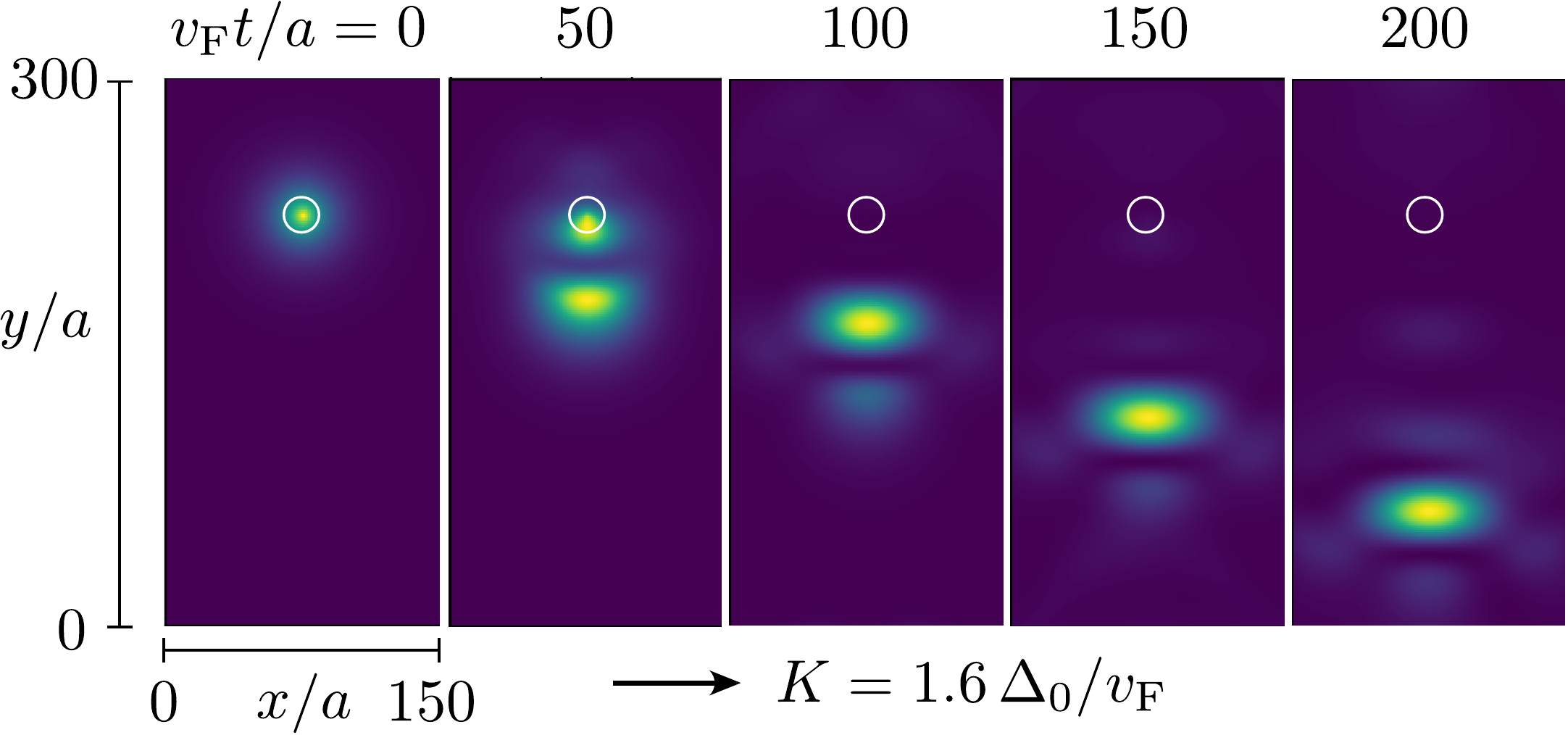}}
\caption{Majorana fermion wave packet that escapes from a vortex (white circle) in response to a quench of the superflow momentum $K$ (in the $x$-direction). The images show a color scale plot of $|\Psi(x,y,t)|^2$ in the plane of the superconductor, at different times following the quench at $t=0$. This is a numerical solution of the evolution equation \eqref{dPsidt}, with Hamiltonian \eqref{calHdef} discretized on a square lattice (lattice constant $a$, $\Delta_0=0.04\,\hbar v_{\rm F}/a$, $B=(h/e)(302\,a)^{-2}$, $\mu=0$). The initial condition at $t=0$ is the Majorana zero-mode $\Psi_+$. A movie of the wave packet propagation is available in the supplementary material. 
}
\label{fig_snapshots}
\end{figure}

The effect of a superflow on a topological superconductor has been demonstrated experimentally \cite{Zhu21} at the proximitized surface of a topological insulator \cite{Fu08}. We focus on that platform \cite{Has10,Qi11}, see Fig.\ \ref{fig_layout}, described by the four-band Bogoliubov-De Gennes Hamiltonian
\begin{align}
{\cal H}_0={}& v_{\rm F}(k_x\sigma_x+k_y\sigma_y)\nu_z-ev_{\rm F}(A_x\sigma_x+A_y\sigma_y)\nu_0\nonumber\\
&-\mu\sigma_0\nu_z+\Delta\sigma_0(\nu_x\cos\phi-\nu_y\sin\phi).\label{H8def}
\end{align}
The surface is in the $x$--$y$ plane, the in-plane momentum is $\bm{k}=-i\partial_{\bm{r}}$. The electron charge is taken as $+e$, the Fermi velocity is $v_{\rm F}$ and $\hbar$ is set to unity. The Pauli matrices $\sigma_\alpha,\nu_\alpha$ act, respectively, on the spin and particle-hole degree of freedom. The corresponding $2\times 2$ unit matrices are $\sigma_0$ and $\nu_0$. An $s$-wave superconducting pair potential $\Delta e^{i\phi}$ couples electrons and holes. Time-reversal symmetry is broken by a perpendicular magnetic field $B$, with vector potential $\bm{A}$.

Charge-conjugation symmetry ${\cal C}=\sigma_y\nu_y{\cal K}$ is expressed by
\begin{equation}
{\cal C}{\cal H}_0{\cal C}=\sigma_y\nu_y {\cal H}_0^\ast\sigma_y\nu_y=-{\cal H}_0.\label{CHC}
\end{equation}
The complex conjugation operation ${\cal K}$ is taken in the real-space basis, so the momentum changes sign. When the Fermi energy $\mu=0$ is at the Dirac point there is additionally a chiral symmetry,
\begin{equation}
\sigma_z\nu_z{\cal H}_0\sigma_z\nu_z=-{\cal H}_0.\label{chiralsym}
\end{equation}

We consider one $h/2e$ vortex at the origin of the coordinate system and in this analysis ignore the presence of other vortices. (The full vortex lattice is included in the computer simulations.) The gap $\Delta$ increases from 0 at the vortex core to $\Delta_0$ outside, on the scale of the superconducting coherence length $\xi_0=\hbar v_{\rm F}/\Delta_0$. The superconducting phase $\phi(\bm{r})$ winds by $\pm 2\pi$ around the vortex, $e^{i\phi(\bm{r})}=r^{-1}(x\pm iy)$. In a strong type-II superconductor ($\xi_0$ much less than the London penetration length) the magnetic field is approximately uniform. We take the gauge where $\bm{A}=-By\hat{x}$. 

The vortex contains a Majorana zero-mode, a charge neutral bound state with zero excitation energy \cite{Fu08}. Its wave function $\Psi$ is an eigenstate of the charge conjugation operator ${\cal C}$. For $\mu=0$ chiral symmetry demands that $\Psi$ is also an eigenstate of $\sigma_z\nu_z$. The combination of the two symmetries enforces the form
\begin{equation}
\begin{split}
&\Psi_+=(e^{i\gamma}\psi_+,0,0,e^{-i\gamma}\psi_+),\\
&\Psi_-=(0,e^{i\gamma}\psi_-,e^{-i\gamma}\psi_-,0),
\end{split}\label{MZMequation}
\end{equation}
for a phase shift $\gamma$ and a pair of real scalar functions $\psi_\pm(\bm{r})$. The sign of the vorticity selects either $\Psi_+$ or $\Psi_-$. An explicit solution \cite{Fu08,Jac81} of ${\cal H}_0\Psi=0$ gives $\gamma=\pi/4$ and an exponential decay of $\psi_\pm$ on the scale of $\xi_0$.

The gapped surface is brought out of equilibrium by a superflow momentum quench $\bm{K}(t)$. The superflow exerts a Magnus force on the Majorana zero mode, which may cause it to escape from the vortex core \cite{Pac21}. A computer simulation of the escape is shown in Fig.\ \ref{fig_snapshots}.

The superflow momentum quench enters the Hamiltonian in the form
\begin{equation}
{\cal H}={\cal H}_0-\bm{K}\cdot\frac{\partial{\cal H}_0}{e\partial \bm{A}}={\cal H}_0+v_{\rm F}(\bm{K}\cdot\bm{\sigma})\nu_0,\label{calHdef}
\end{equation}
in accord with Galilean invariance. We assume an instantaneous quench in the $x$-direction, $\bm{K}(t)=K\theta(t)\hat{x}$, so we seek the solution of the evolution equation
\begin{equation}
i\partial_t\Psi(t)=({\cal H}_0+v_{\rm F}K\sigma_x\nu_0)\Psi(t),\label{dPsidt}
\end{equation}
with initial condition $\Psi(0)=\Psi_\pm$ given by Eq.\ \eqref{MZMequation} The quench preserves both particle-hole and chiral symmetries.

The full superflow momentum
\begin{equation}
\bm{P}(\bm{r})=\bm{p}_s(\bm{r})+\bm{K}
\end{equation}
 includes also the contribution from the circulating momentum field $\bm{p}_s$ around the vortex cores. This divergence-free field has the gauge invariant expression \cite{Tinkham}
\begin{equation}
\bm{p}_s(\bm{r})=\tfrac{1}{2}\nabla\phi(\bm{r})-e\bm{A}(\bm{r}).
\end{equation}

For later use we note that the gauge transformation 
\begin{align}
{\cal H}\mapsto{}& e^{-i\phi(\bm{r})\nu_z/2}{\cal H}e^{i\phi(\bm{r})\nu_z/2}\nonumber\\
={}& v_{\rm F}(\bm{k}\cdot\bm{\sigma})\nu_z+v_{\rm F}(\bm{P}\cdot\bm{\sigma})\nu_0-\mu\sigma_0\nu_z+\Delta\sigma_0\nu_x\label{Hvs}
\end{align}
explicitly writes the Hamiltonian in terms of the full superflow momentum. 

\section{Short-time dynamics}
\label{sec_short}

For the initial time dependence we may truncate the Taylor expansion of the propagator $e^{-it{\cal H}}$,
\begin{equation}
\Psi(t)=e^{-it{\cal H}}\Psi_\pm=\sum_{n=0}^\infty \frac{(-it)^n}{n!}{\cal H}^n\Psi_\pm.
\end{equation}
We calculate $\Psi(t)$ to order $t^4$, with the help of the polar-coordinate identity
\begin{equation}
(k_x\pm ik_y)f(r,\varphi)=\mp e^{\pm i\varphi}\left(\pm i \frac{\partial f}{\partial r}-r^{-1}\frac{\partial f}{\partial\varphi}\right),
\end{equation}
and then compute the expectation value of the velocity,
\begin{equation}
\langle \dot{r}_\alpha(t)\rangle=\langle\Psi(t)|\partial{\cal H}/\partial k_\alpha|\Psi(t)\rangle=v_{\rm F}\langle\Psi(t)|\sigma_\alpha\nu_z|\Psi(t)\rangle.
\end{equation}
We focus on the case $\mu=0$ of chiral symmetry.

To simplify the calculation we note that the magnetic field only affects the dynamics on the scale of the magnetic length $l_m=\sqrt{\hbar/eB}$, which is large compared to the vortex size $\xi_0$ for magnetic fields small compared to the upper critical field of the superconductor. For the short-time dynamics we may ignore the magnetic field. In terms of the gap profile $\Delta(r)$ the scalar function $\psi_\pm$ in the initial state \eqref{MZMequation} is then given by \cite{Fu08,Jac81} 
\begin{equation}
\psi_+(r)=\psi_-(r)=c\exp\left(-v_{\rm F}^{-1}\int_0^r dr'\Delta(r')\right),
\end{equation} 
with $c$ a normalization constant.

A simple closed-form expression results for a constant $\Delta\equiv\Delta_0$,
\begin{subequations}
\label{xdotydot}
\begin{align}
\langle \dot{x}(t)\rangle={}&-2 v_{\rm F}^2\Delta_0^2 K t^3 \cos 2 \gamma +{\cal O}(t^5),\\
\langle \dot{y}(t)\rangle={}&-2v_{\rm F}^2K t+\tfrac{4}{3}v_{\rm F}^4K^3 t^3 \nonumber\\
&+\tfrac{2}{3}v_{\rm F}^2\Delta_0^2K t^3(10  - 9 \sin 2 \gamma) +{\cal O}(t^5).
\end{align}
\end{subequations}
These are the formulas for $+2\pi$ vorticity (initial condition $\Psi_+$); for $-2\pi$ vorticity (initial condition $\Psi_-$) the component $\langle \dot{x}\rangle$ is unchanged while $\langle\dot{y}\rangle$ changes sign.

The zero-mode has $\gamma=\pi/4$, hence the motion is fully in the $y$-direction, with initial velocity
\begin{equation}
\langle \dot{y}(t)\rangle=\pm 2v_{\rm F}^2K t\bigl(-1+\tfrac{1}{3}t^2(\Delta_0^2+2v_{\rm F}^2K^2) +{\cal O}(t^4)\bigr),\label{dotypi4}
\end{equation}
for $\pm 2\pi$ vorticity. Because of the dependence on the vorticity, we interpret the initial acceleration $\pm 2v_{\rm F}^2K$ as a manifestation of the Magnus force acting on the zero-mode.

One may wonder whether the Lorentz force, which we have ignored in this calculation, would deflect the particle away from the $y$-axis. This is not the case, chiral symmetry enforces $\langle\dot{x}(t)\rangle=0$ for all $t>0$ when $\gamma=\pi/4$, see App.\ \ref{app_Lorentzforce}.

\section{Semiclassical calculation of the escape velocity}
\label{sec_semiclassics}

A semiclassical approximation will allow us to obtain a simple analytical expression for the velocity at which the Majorana fermion escapes from the vortex core. For simplicity, we set $\mu=0$, so chiral symmetry applies.

Quite generally, a wave packet at position $\bm{r}$ and with wave vector $\bm{k}$ has the semiclassical equations of motion \cite{Xia10,Wan21}
\begin{subequations}
\label{eq_EOM}
 \begin{align}
&\dot{\bm{r}}=\partial_{\bm{k}}E-\dot{\bm{k}}\times(\partial_{\bm{k}}\times\mathbf{\cal A})+(\dot{\bm{r}}\cdot\partial_{\bm{r}})\mathbf{\cal A}-\partial_{\bm{k}}(\bm{a}\cdot\dot{\bm{r}})\\
&\dot{\bm{k}}=-\partial_{\bm{r}}E+\dot{\bm r}\times(\partial_{\bm{r}}\times \bm{a})-(\dot{\bm{k}}\cdot\partial_{\bm{k}})\bm{a}+\partial_{\bm{r}}(\mathbf{\cal A}\cdot\dot{\bm{k}}).
\end{align}
\end{subequations}
The energy $E(\bm{r},\bm{k})$ is an eigenvalue of the $4\times 4$ matrix $H(\bm{r},\bm{k})$, obtained from the Hamiltonian ${\cal H}$ by treating $\bm{r}$ and $\bm{k}$ as parameters --- not as operators. The corresponding eigenfunction $|u(\bm{r},\bm{k})\rangle$ is a rank-4 spinor, normalized to unity, $\langle u|u\rangle=1$. 

The fields $\mathbf{\cal A}$ and $\bm{a}$ are defined by the connections
\begin{subequations}
\label{eq_connections}
\begin{align}
&\mathbf{\cal A}(\bm{r},\bm{k})=\langle u(\bm{r},\bm{k})|i\partial_{\bm{k}}|u(\bm{r},\bm{k})\rangle,\\
&\bm{a}(\bm{r},\bm{k})=\langle u(\bm{r},\bm{k})|i\partial_{\bm{r}}|u(\bm{r},\bm{k})\rangle.
\end{align}
\end{subequations}
The state $|u\rangle$ is defined up to a complex phase factor. If $|u\rangle\mapsto e^{if(\bm{r},\bm{k})}|u\rangle$ the connections transform as $\mathbf{\cal A}\mapsto \mathbf{\cal A}-\partial_{\bm{k}}f$ and $\bm{a}\mapsto\bm{a}-\partial_{\bm{r}}f$. These two transformations leave the right-hand-side of Eq.\ \eqref{eq_EOM} unchanged.

We apply this general formalism to the Hamiltonian \eqref{Hvs}, to ensure that the full gauge invariant superflow momentum appears in the equations of motion. Diagonalization of $H(\bm{r},\bm{k})$ for $\mu=0$ gives four eigenstates $|u_n(\bm{r},\bm{k})\rangle$ with eigenvalues
\begin{widetext}
\begin{equation}
\begin{split}
&E_n(\bm{r},\bm{k})= s_n\sqrt{v_{\rm F}^2 P(\bm{r})^2+\Delta(\bm{r})^2+v_{\rm F}^2 k^2+2s'_n v_{\rm F}\sqrt{P(\bm{r})^2\Delta(\bm{r})^2+v_{\rm F}^2\bigl(\bm{k}\cdot\bm{P}(\bm{r})\bigr)^2}},\\
&\{s_1^{\vphantom{'}},s'_1\}=\{+,+\},\;\;\{s_2^{\vphantom{'}},s'_2\}=\{-,+\},\;\;\{s_3^{\vphantom{'}},s'_3\}=\{+,-\},\;\;\{s_4^{\vphantom{'}},s'_4\}=\{-,-\}.
\end{split}
\end{equation}
\end{widetext} 

We find that the connections \eqref{eq_connections} do not contribute to the equations of motion \eqref{eq_EOM}, because they are given by the gradient of a scalar field,
\begin{align}
&\mathbf{\cal A}_n(\bm{r},\bm{k})=\partial_{\bm{k}}f_n(\bm{r},\bm{k}),\;\;\bm{a}_n(\bm{r},\bm{k})=\partial_{\bm{r}}f_n(\bm{r},\bm{k}),\\
&f_n=\tfrac{1}{2} \arctan\left(\frac{P^2+s'_n\sqrt{(\bm{k}\cdot\bm{P})^2+P^2 \Delta^2/v_{\rm F}^2}}{k_x P_y-k_y P_x}\right)\nonumber\\
&\qquad-\tfrac{1}{2} \arctan(P_y/P_x).
\end{align}
The semiclassical dynamics is therefore fully determined by the energy landscape,
\begin{equation}
\dot{\bm{r}}=\partial_{\bm{k}}E_n,\;\;\dot{\bm{k}}=-\partial_{\bm{r}}E_n.
\end{equation}

We consider a rotationally symmetric vortex, $\Delta(\bm{r})=\Delta(r)$, $\bm{P}(\bm{r})=\bm{K}\hat{x}+p_s(r)\hat{z}\times\hat{\bm{r}}$, so that $E(-x,y,-k_x,k_y)=E(x,y,k_x,k_y)$. A trajectory that starts out with $x=0$, $k_x=0$ at $t=0$ then will retain these values for $t>0$. The motion along the $y$-axis is determined by the equations of motion
\begin{subequations}
\begin{align}
\dot{y}={}&\partial_{k_y} E_n=v_{\rm F}^2 k_y/E_n,\\
\dot{k}_y={}&-\partial_y E_n=-\frac{1}{2E_n}\frac{d}{dy}\bigl[s'_nv_{\rm F}P(y)+ \Delta(y)\bigr]^2.
\end{align}
\end{subequations}
We denote $\Delta(x=0,y)\equiv \Delta(y)$ and $P(x=0,y)\equiv P(y)$. 

Since $dE_n/dt=0$, the energy $E_n$ is equal to its value at $t=0$. Assuming $\Delta(0)=0$ and $P(0)=|K|$ (vanishing pair potential and no circulating superflow deep inside the vortex core), we have $E_n=s_nv_{\rm F}|K|$.

Far outside of the core, where $\Delta\approx\Delta_0$ and $P\approx |K|$ are both $y$-independent, one has $\dot{k}_y=0$. The terminal $k_y$ should satisfy
\begin{equation}
v_{\rm F}^2 K^2=E_n^2\Rightarrow
 v_{\rm F}^2 k_y^2=-\Delta_0^2-2s'_n v_{\rm F}|K|\Delta_0,
\end{equation}
which has a real solution for $s'_n=-1$ if $v_{\rm F}|K|>\Delta_0/2$. That is the condition for escape of the Majorana fermion. The escape velocity is given by
\begin{equation}
|v_{\rm escape}|=|K|^{-1}\sqrt{2v_{\rm F}|K|\Delta_0-\Delta_0^2}.\label{vescape}
\end{equation}
The maximum $|v_{\rm escape}|=v_{\rm F}$ is reached at $|K|=\Delta_0/v_{\rm F}$.

Notice that the quenched superconductor supports a quasiparticle escape even though the excitation gap has not closed: the reduced gap $\Delta_{\rm eff}=\Delta_0-v_{\rm F}|K|$ only closes for $v_{\rm F}|K|>\Delta_0$, while quasiparticle escape is possible for $v_{\rm F}|K|>\Delta_0/2$. The reason is that the quench gives an excess energy $v_{\rm F}|K|$ to the quasiparticle, so escape becomes possible when $v_{\rm F}|K|>\Delta_{\rm eff}\Rightarrow v_{\rm F}|K|>\Delta_0/2$. The quasiparticle will eventually loose its excess energy by inelastic processes, which are not included in our calculations.

\begin{figure}[tb]
\centerline{\includegraphics[width=0.8\linewidth]{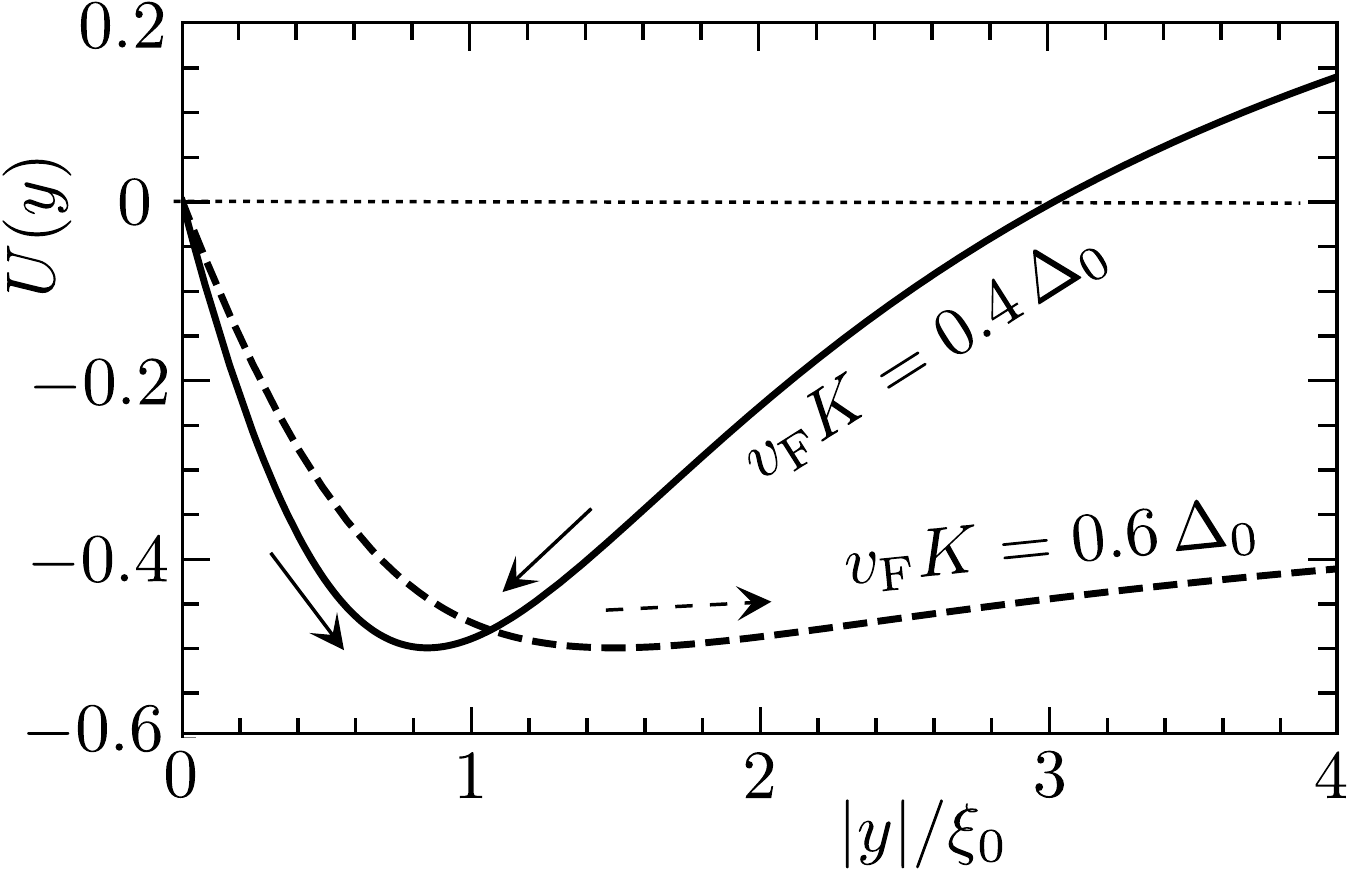}}
\caption{Plot of the potential $U(y)$ that governs the equation of motion \eqref{dotydotky} of the wave packet, calculated for the gap and superflow velocity profiles \eqref{deltapprofile}. The arrows indicate the oscillatory motion for $v_{\rm F}|K|<\Delta_0/2$ and the escape to infinity for $v_{\rm F}|K|>\Delta_0/2$. The direction in which the wave packet escapes is minus the sign of $K$ times the sign of the vorticity.
}
\label{fig_Uplot}
\end{figure}

The escaping wave packet is a superposition of the two states $|u_n\rangle$ with $s'_n=-1$ and $s_n=\pm 1$. They satisfy the same equation of motion
\begin{equation}
\begin{split}
&\ddot{y}=-v_{\rm F}^2U'(y),\\
&U(y)=\frac{1}{2K^2}\left(\bigl[ P(y)-\Delta(y)/v_{\rm F}\bigr]^2-K^2\right),
\end{split}\label{dotydotky}
\end{equation}
with initial conditions $\dot{y}(0)=0$ and $y(0)$ infinitesimal (needed to avoid the discontinuous derivative $\Delta'(y)$ at $y=0$)\cite{note1}. This is the frictionless motion in the potential landscape $U(y)$, plotted in Fig.\ \ref{fig_Uplot} for the functional forms
\begin{equation}
\Delta(y)=\frac{\Delta_0 |y|}{\sqrt{y^2+\xi_0^2}},\;\;P(y)=|K|+\frac{|y|/2}{y^2+\xi_0^2}\label{deltapprofile}
\end{equation}
appropriate for a vortex with coherence length $\xi_0$ much smaller than the London penetration length \cite{Tinkham,Cle75}.

For $v_{\rm F}|K|>\Delta_0/2$ one has $U(\infty)<U(0)$ so the motion escapes to infinity, with a constant terminal velocity \eqref{vescape}, for $v_{\rm F}|K|<\Delta_0/2$ the motion is oscillatory.

The two states $s_n=\pm 1$, at energies $\pm v_{\rm F}|K|$, are related by particle-hole symmetry, they have the same position but opposite momentum. The semiclassical calculation neglects interference of the positive and negative energy states, which is reliable for the long-time dynamics outside of the vortex core, when the momentum difference is large and interference effects average out. In contrast, inside the vortex core the two states both still have momentum approximately equal to zero, and their interference cannot be neglected.

\section{Computer simulations}
\label{sec_simulations}

We have simulated the wave packet dynamics by discretizing the Bogoliubov-De Gennes Hamiltonian \eqref{H8def} on a square lattice (lattice constant $a$),
\begin{subequations}
\label{H0discrete}
\begin{align}
H={}& (v_{\rm F}/a)(\sigma_x\sin ak_x+\sigma_y\sin ak_y)\nu_z-\mu\sigma_0\nu_z\nonumber\\
&+M(k)\sigma_z\nu_0-ev_{\rm F}(A_x\sigma_x+A_y\sigma_y)\nu_0\nonumber\\
&+\Delta\sigma_0(\nu_x\cos\phi-\nu_y\sin\phi)+v_{\rm F}K\sigma_x\nu_0,\\
M(k)={}&M_0-(M_1/a^2)(2-\cos ak_x-\cos ak_y),
\end{align}
\end{subequations}
and evolving the zero-mode wave function via a finite-difference algorithm. The $M(k)$ term in Eq.\ \eqref{H0discrete} includes the effect of a small coupling between the top and bottom surfaces of the topological insulator of Fig.\ \ref{fig_layout}. As in Ref.\ \onlinecite{Pac21} we set $M_0=0$, to avoid the opening of a gap at $\bm{k}=0$, but retain a nonzero $M_1 = 0.2\,av_{\rm F}$ in order to eliminate the fermion doubling at $\bm{k} = (\pi/a, \pi/a)$. 

We take a uniform magnetic field $B\hat{z}=\nabla\times\bm{A}$, appropriate for a strong type-II superconductor. The vortex array has a pair of $h/2e$ vortices in a magnetic unit cell of size $d_0\times d_0$, with $d_0=302\,a$ (corresponding to a magnetic field $B=h/ed_0^{2}$.) The phase field $\phi(\bm{r})$ winds by $2\pi$ around each vortex, at position $\bm{R}_n$, as expressed by
\begin{equation}
\nabla\times\nabla \phi(\bm{r})= 2\pi\hat{z}\textstyle{\sum_{n}}\delta(\bm{r}-\bm{R}_n),\;\;\nabla^2\phi=0.\label{phidef}
\end{equation}

For the pair potential in a vortex core we take the gap profile $\Delta(r)=\Delta_0\tanh(r/r_0)$, with $\Delta_0=0.2\, v_{\rm F}/a$. The core size $r_0$ is of order $\xi_0=v_{\rm F}/\Delta_0$, but for the sake of comparison with the semiclassics (which assumes a smooth gap profile) we will also consider larger values of $r_0$. The gap $\Delta(r)$ is saturated at $\Delta_0$ for $r>70\,a$, to ensure that the vortex core is fully contained within a single magnetic unit cell. We follow the dynamics of the wave packet on a time scale that is sufficiently short that only a single vortex plays a dominant role. To avoid interference from the other vortex we set its core size to zero. 

We use the package Tkwant for the calculations \cite{Klo21,zenodo}. See App.\ \ref{app_simulations} for details on the simulation.

\begin{figure}[tb]
\centerline{\includegraphics[width=1\linewidth]{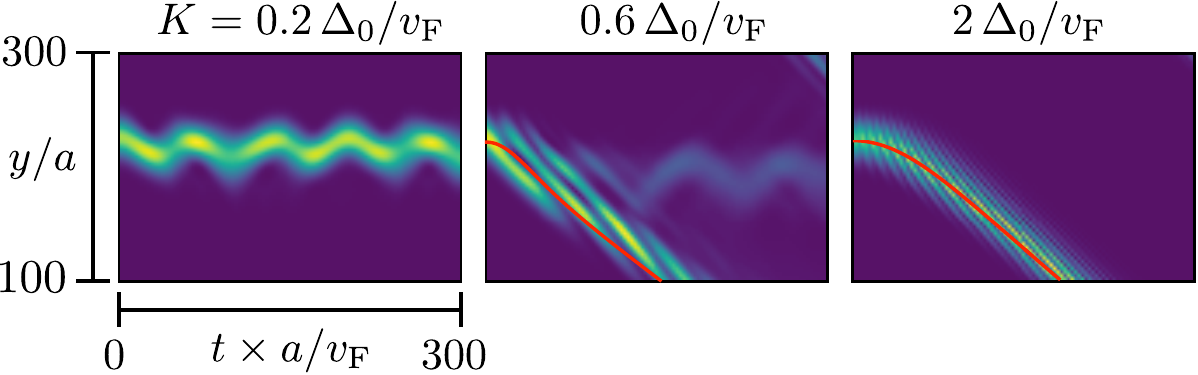}}
\caption{Propagation of the zero-mode along the $y$-axis after the superflow momentum quench at $t=0$. The color scale shows the density profile $\int |\Psi(x,y,t)|^2\,dx$ following from the computer simulation ($\Delta_0=0.2\,v_{\rm F}/a$, $r_0=40\,a$, $B=(h/e)(302\,a)^{-2}$). The red curve results from integration of the semiclassical equation of motion \eqref{dotydotky}, for the same $\Delta(r)=\Delta_0\tanh(r/r_0)$ gap profile as in the numerics \cite{note2}.  
}
\label{fig_colorscale}
\end{figure}

In Fig.\ \ref{fig_colorscale} we show the time dependence of the propagation of the wave packet along the $y$-axis, following a superflow quench at $t=0$. We compare $\int |\Psi(x,y,t)|^2\,dx$ from the simulation with $y(t)$ from the semiclassical equation of motion \eqref{dotydotky}. The comparison has no adjustable parameters. As anticipated, the agreement is good outside of the vortex core ($K\gtrsim\Delta_0/v_{\rm F})$, where the interference of the positive and negative energy wave packets can be neglected. The oscillatory motion of the wave packet inside the vortex, for small $K$, is not well described by the semiclassics.

\begin{figure}[tb]
\centerline{\includegraphics[width=0.8\linewidth]{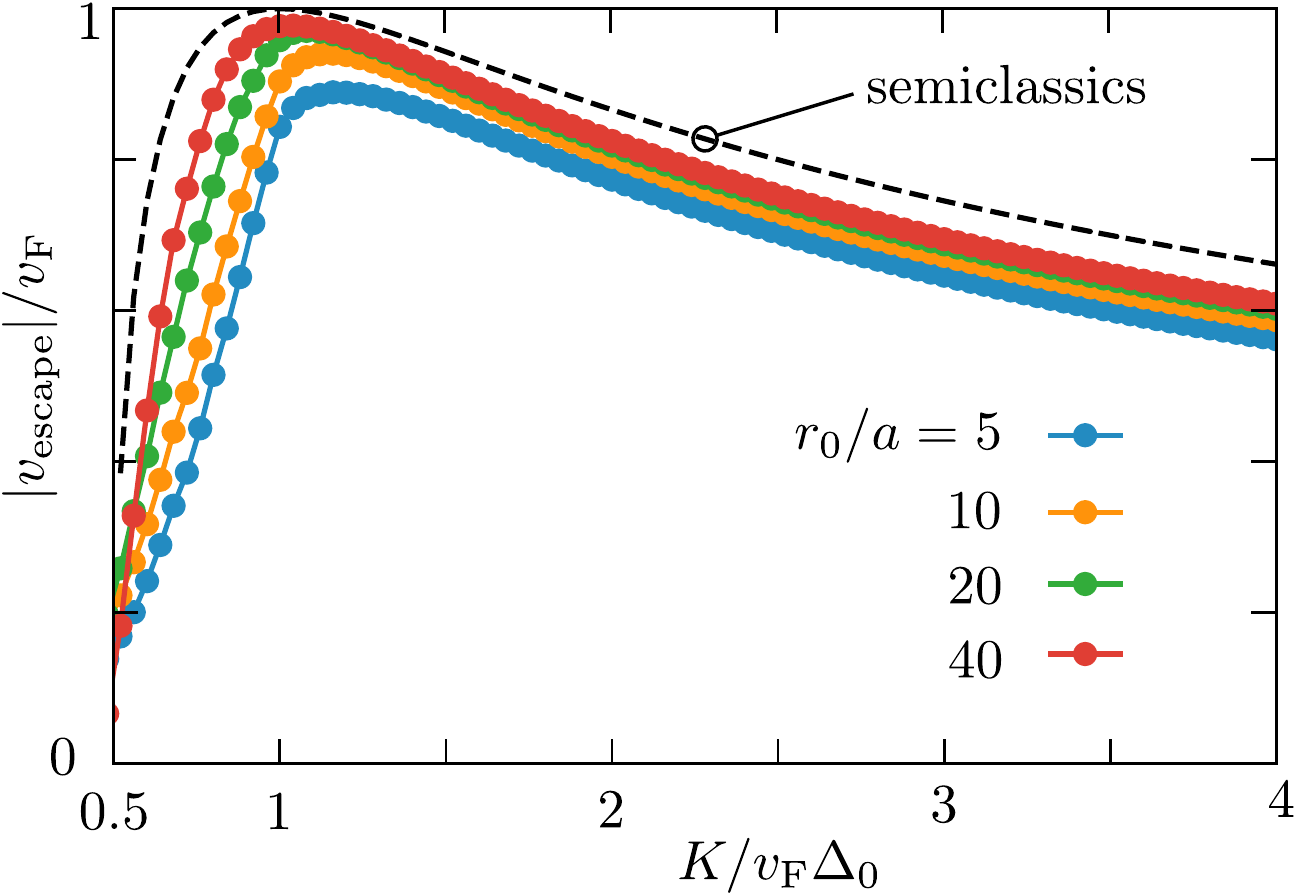}}
\caption{Escape velocity of a zero-mode wave packet from the vortex core, as a function of the superflow momentum $K$. The data points follow from the computer simulation, for different core sizes $r_0$ (at fixed $\Delta_0=0.2\,v_{\rm F}/a$). The black dashed curve is the semiclassical result \eqref{vescape}.
}
\label{fig_escape}
\end{figure}

Fig.\ \ref{fig_escape} compares the escape velocity obtained from the simulation with the semiclassical formula \eqref{vescape}. The numerical data nicely approaches the semiclassics for larger and larger core sizes.

\section{Conclusion}
\label{sec_conclude}

In summary, we have investigated the dynamics of the Majorana delocalization transition reported in Ref.\ \onlinecite{Pac21}. A supercurrent can be used to extract a Majorana fermion from the zero-mode bound to a vortex core. The extraction process is governed by an effective potential well, see Fig.\ \ref{fig_Uplot}, which allows for escape with a constant terminal velocity $v_{\rm escape}$ once the supercurrent exceeds a critical value. A simple semiclassical calculation of this velocity agrees well with computer simulations.

The escape of the Majorana fermion should be observable by scanning probe spectroscopy, as a current pulse when the probe is positioned near a vortex, at right angles from the superflow. Close to the deconfinement transition the escape velocity will be much smaller than the Fermi velocity $v_{\rm F}\approx 10^5\,\text{m/s}$ (see Fig.\ \ref{fig_escape}), which should make the observation more feasible.

The internal degree of freedom of the Majorana zero-mode that couples to the superflow via the Magnus effect is the chirality --- zero-modes of opposite chirality escape from the vortex in opposite directions. The conformal field theory of non-Abelian anyons associates a ``topological spin'' to a Majorana zero-mode \cite{Nay08,Tu13,Ari17}. As a topic for future research we ask whether there is an analogous Magnus effect for the topological spin. We note that the phase shift $\gamma$ in the Majorana wave function \eqref{MZMequation} affects the direction in which the superflow drives the quasiparticle, see Eq.\ \eqref{xdotydot}. The motion is strictly perpendicular to the superflow only for $\gamma=\pi/4$. That this also happens to be the value of the topological spin may or may not be accidental.

\acknowledgments	
We thank F. Hassler for helpful discussions. This project has received funding from the Netherlands Organization for Scientific Research (NWO/OCW) and from the European Research Council (ERC) under the European Union's Horizon 2020 research and innovation programme.

\appendix

\section{Chiral symmetry prevents lateral deflection by the Lorentz force}
\label{app_Lorentzforce}

Fig.\ \ref{fig_snapshots} shows that the Majorana fermion escapes from the vortex along the $y$-direction, perpendicular to the superflow. We address the question why the motion is not bent in the $x$-direction by the Lorentz force. Since electrons and holes are deflected in \textit{the same} direction by the Lorentz force, charge-neutrality of the quasiparticle does not prevent the deflection. Chiral symmetry is essential.

To demonstrate this, we calculate the expectation value at $\mu=0$ of the $x$-component of the velocity operator,
\begin{equation}
\langle \dot{x}(t)\rangle=v_{\rm F}\langle\Psi(0)|e^{i{\cal H}t}\sigma_x\nu_ze^{-i{\cal H}t}|\Psi(0)\rangle.\label{vxaverage}
\end{equation}

The superconducting vortex at the origin has pair potential $\Delta(r) e^{\pm i\phi}$, in polar coordinates $(r,\phi)$, with a rotationally symmetric amplitude $\Delta(r)$ and a $\pm 2\pi$ vorticity. The magnetic field $B(r)\hat{z}$ is also assumed to be rotationally symmetric, with vector potential $\bm{A}(\bm{r})=g(r)(-y,x,0)$ [so that $B(r)=2g(r)+rg'(r)$].

The initial state $\Psi(0)=\Psi_\pm$ is a zero-mode bound to the vortex core, given by \cite{Fu08,Jac81}
\begin{equation}
\begin{split}
&\Psi_+=c_+e^{+\chi(\bm{r})}\exp\left(-v_{\rm F}^{-1}\int_0^r \Delta(r')dr'\right)
\small{
\begin{pmatrix}
e^{i\pi/4}\\
0\\
0\\
e^{-i\pi/4}
\end{pmatrix},
}
\\
&\Psi_-=c_-e^{-\chi(\bm{r})}\exp\left(-v_{\rm F}^{-1}\int_0^r \Delta(r')dr'\right)
\small{
\begin{pmatrix}
0\\
e^{i\pi/4}\\
e^{-i\pi/4}\\
0
\end{pmatrix},
}
\end{split}\label{Psipmapp}
\end{equation}
with $c_\pm $ a normalization constant and $\chi(\bm{r})$ chosen such that 
\begin{equation}
\partial_y\chi=eA_x,\;\;\partial_x\chi=-eA_y.
\end{equation}
Note that $\chi(-x,y)=\chi(x,y)$.

We introduce the operator ${\cal P}_x$ which reflects $x\mapsto-x$, $k_x\mapsto-k_x$. Its action on the Hamiltonian ${\cal H}={\cal H}_0+v_{\rm F}K\sigma_x\nu_0$ is given by
\begin{equation}
{\cal P}_x{\cal H}{\cal P}_x=\sigma_x\nu_y{\cal H}\sigma_x\nu_y\;\;\text{if}\;\;\mu=0,
\end{equation}
see Eq.\ \eqref{H8def}. The zero-mode \eqref{Psipmapp} is unchanged upon reflection, $\Psi_\pm(x,y)=\Psi_\pm(-x,y)$, and moreover
\begin{equation}
\Psi_\pm=-\sigma_x\nu_y\Psi_\pm.
\end{equation}
These identities imply that
\begin{equation}
{\cal P}_xe^{-i{\cal H} t}\Psi_\pm=-\sigma_x\nu_y e^{-i{\cal H} t}\Psi_\pm.
\end{equation}

We now calculate, using also $\sigma_x\nu_z={\cal P}_x\sigma_x\nu_z{\cal P}_x$, the expectation value \eqref{vxaverage},
\begin{align}
\langle \dot{x}(t)\rangle={}&v_{\rm F}\langle\Psi_\pm|e^{i{\cal H}t}\sigma_x\nu_ze^{-i{\cal H}t}|\Psi_\pm\rangle\nonumber\\
={}&v_{\rm F}\langle\Psi_\pm|e^{i{\cal H}t}{\cal P}_x\sigma_x\nu_z{\cal P}_xe^{-i{\cal H}t}|\Psi_\pm\rangle\nonumber\\
={}&v_{\rm F}\langle\Psi_\pm|e^{i{\cal H}t}(-\sigma_x\nu_y)(\sigma_x\nu_z)(-\sigma_x\nu_y)e^{-i{\cal H}t}|\Psi_\pm\rangle\nonumber\\
={}&-v_{\rm F}\langle\Psi_\pm|e^{i{\cal H}t}\sigma_x\nu_ze^{-i{\cal H}t}|\Psi_\pm\rangle\nonumber\\
={}&-\langle \dot{x}(t)\rangle\Rightarrow \langle \dot{x}(t)\rangle\equiv 0.
\end{align}
The velocity component in the $x$-direction has zero expectation value for all $t$, there is no lateral deflection by the Lorentz force at $\mu=0$.

\section{Details of the numerical calculations}
\label{app_simulations}

\begin{figure}[tb]
\centerline{\includegraphics[width=0.8\linewidth]{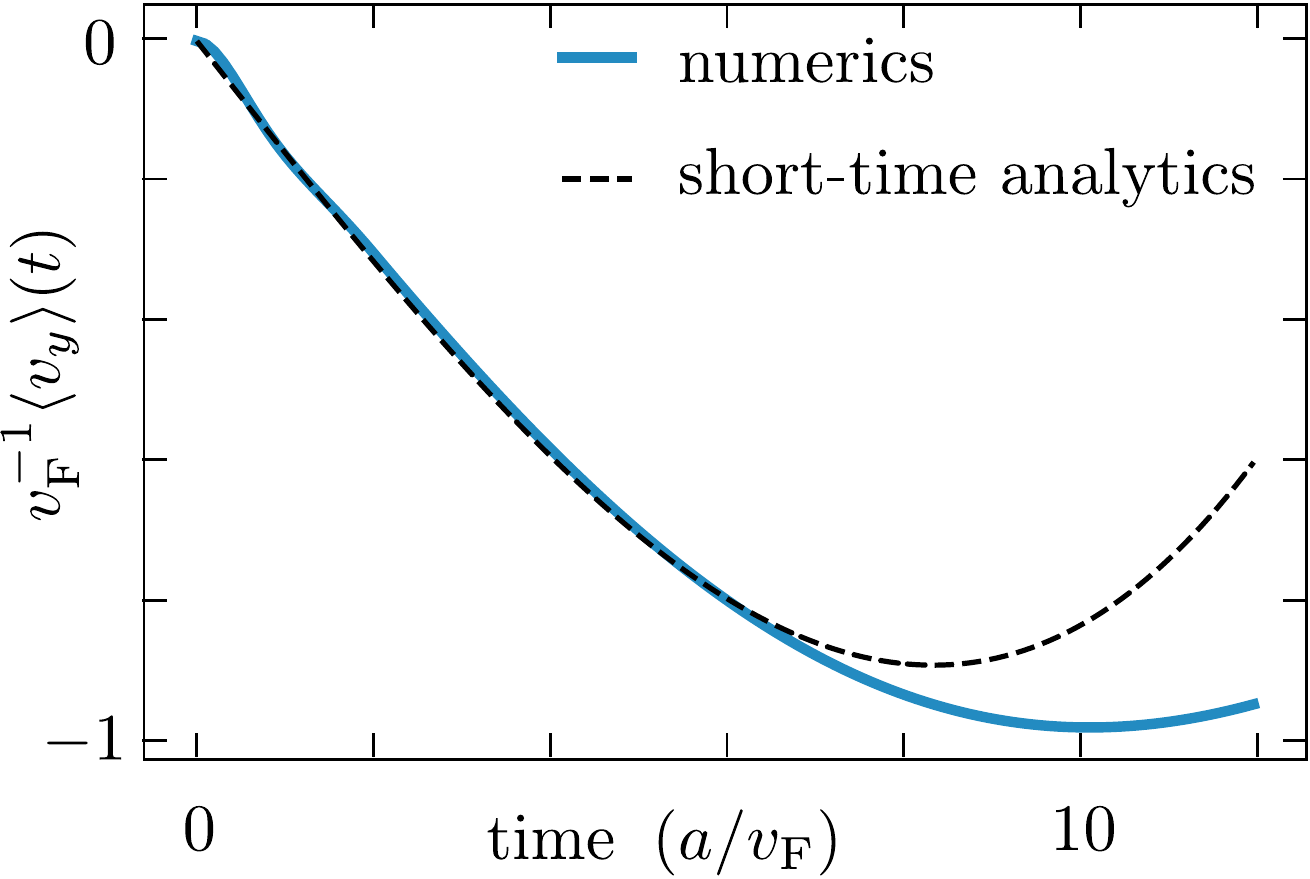}}
\caption{Initial time dependence of the expectation value of the velocity $v_y$ of the Majorana wave packet (perpendicular to the superflow), for $\Delta_0=0.04\,v_{\rm F}/a$, $K=2\Delta_0/v_{\rm F}$, in the limit $r_0\rightarrow 0$ of a small vortex core. The dashed curve is the analytical result from Eq.\ \eqref{dotypi4}.
}
\label{fig_shorttime}
\end{figure}

\begin{figure}[tb]
\centerline{\includegraphics[width=0.9\linewidth]{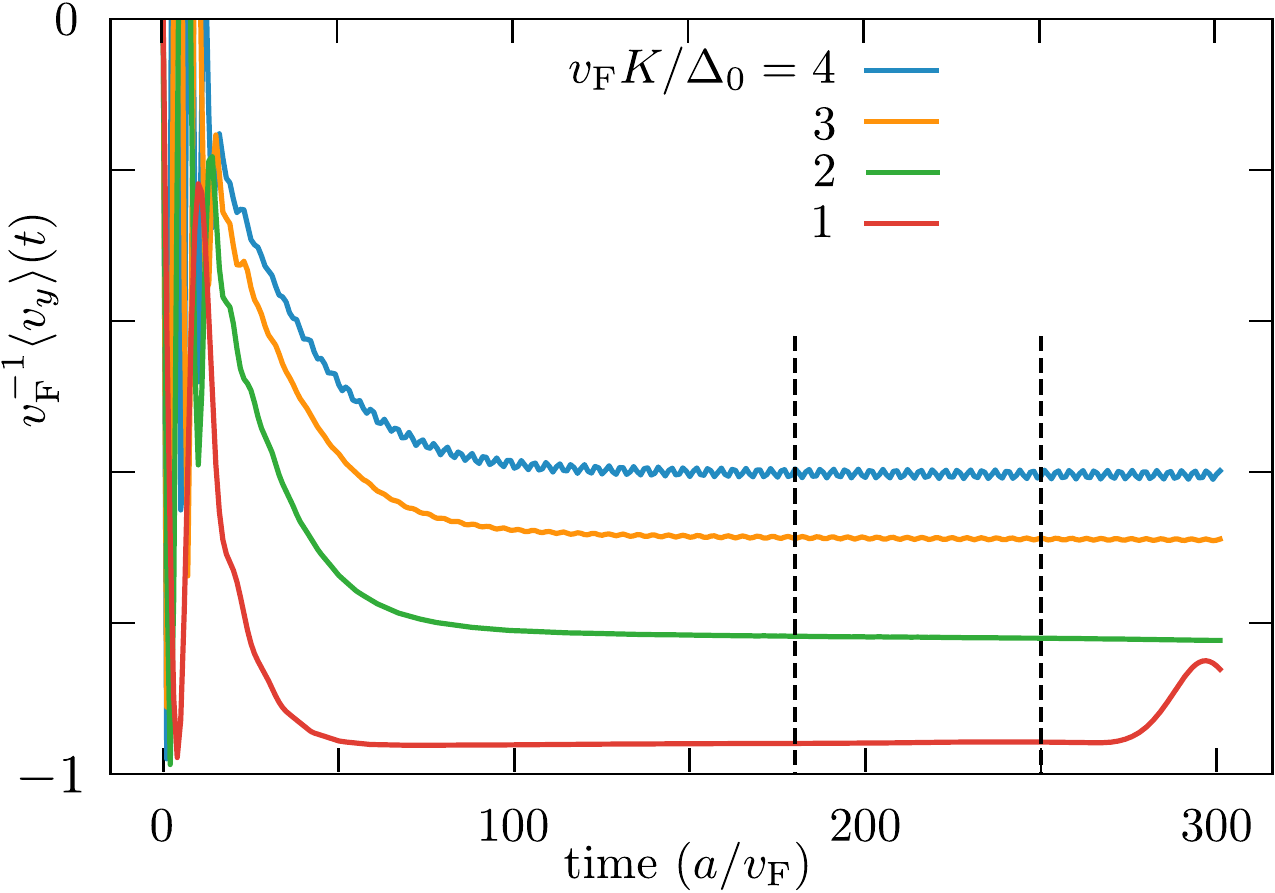}}
\caption{Time dependence of the velocity when the Majorana wave packet is driven out of the vortex core by the Magnus force. Four values of the superflow momentum $K$ are shown at fixed $\Delta_0=0.2\,v_{\rm F}/a$. The average over the time interval between the dashed lines is the escape velocity plotted in Fig.\ \ref{fig_escape} (green curve, for $r_0=20\,a$). For much shorter times the wave packet is still trapped in the vortex core. For longer times the wave packet reaches the boundary of the magnetic unit cell.
}
\label{fig_velocity}
\end{figure}

The velocity operator is given by $\bm{v}=\partial H/\partial \bm{k}$, with $H$ the tight-binding Hamiltonian \eqref{H0discrete}. (In the continuous limit this reduces to $v_i=v_{\rm F}\sigma_i\nu_z$.) We compute the expectation value $\langle v_y\rangle(t)$ as function of time. As a consistency check we show in Fig.\ \ref{fig_shorttime} the short-time dynamics together with the analytical result \eqref{dotypi4}. For longer times the wave packet may escape from the vortex core. We determine the escape velocity by averaging $\langle v_y\rangle(t)$ over a brief time interval, see Fig.\ \ref{fig_velocity}.  

\begin{figure}[tb]
\centerline{\includegraphics[width=0.8\linewidth]{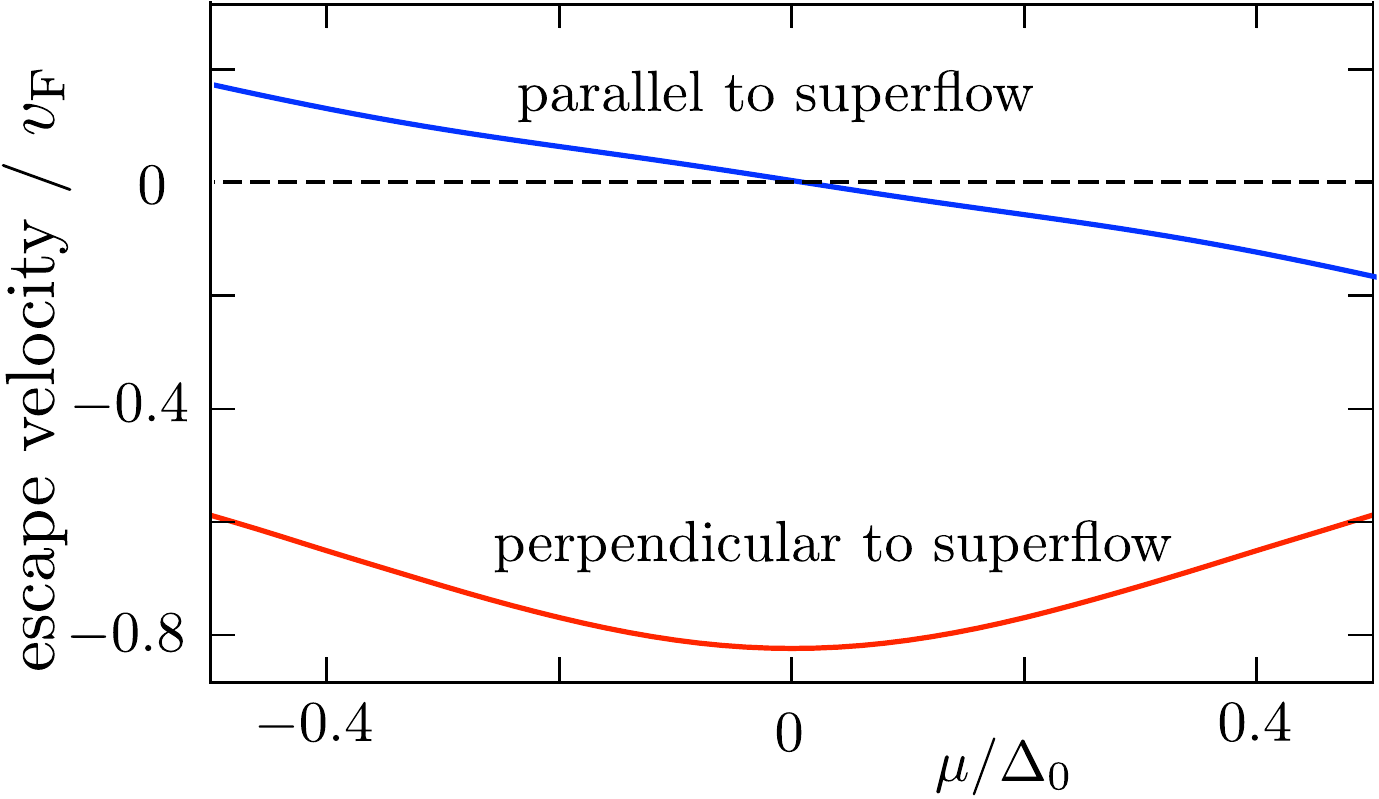}}
\caption{Dependence of the direction of the escape velocity on the chemical potential $\mu$ (for fixed $\Delta_0=0.2\,v_{\rm F}/a$, $K=2\Delta_0/v_{\rm F}$, and vortex core size $r_0=40\,a$).
}
\label{fig_vx}
\end{figure}

This is all data for $\mu=0$, when the expectation value of the velocity component $v_x$ parallel to the superflow vanishes. A nonzero $\mu$ breaks chiral symmetry and introduces a nonzero $\langle v_x\rangle$, see Fig.\ \ref{fig_vx}. The sign of $\mu$ dictates the direction of the deflection away from the $y$-axis.

\begin{figure}[tb]
\centerline{\includegraphics[width=0.9\linewidth]{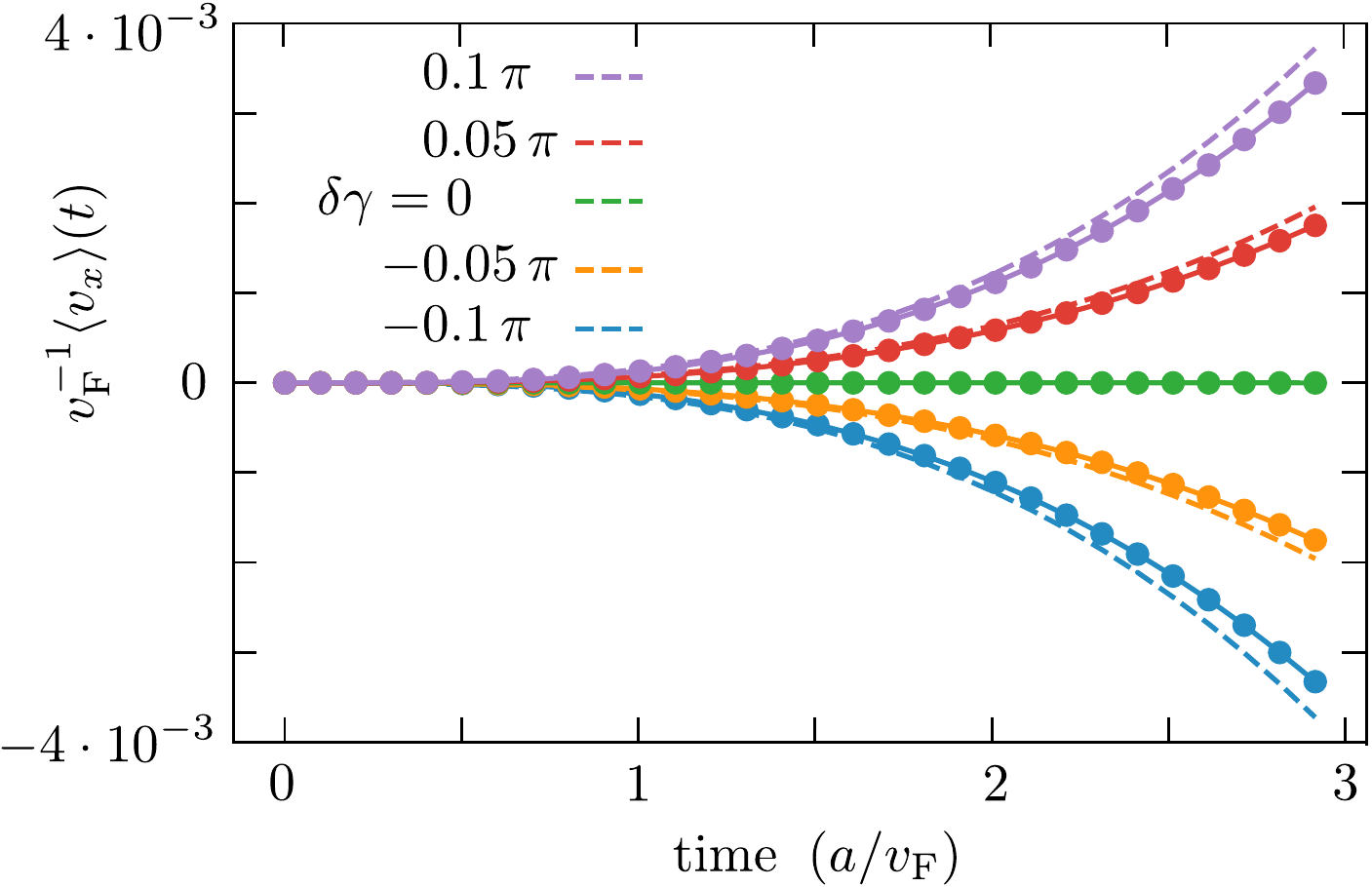}}
\caption{Dependence of the velocity $v_x$ parallel to the superflow on the phase shift $\gamma=\pi/4+\delta\gamma$ between the electron and hole components of the initial wave packet (for $\Delta_0=0.04\,v_{\rm F}/a$, $K=2\Delta_0/v_{\rm F}$, $r_0\rightarrow 0$). This is data for $\mu=0$, the deflection in the $x$-direction happens when $\gamma$ is pushed away from $\pi/4$ by an initial voltage pulse. The solid curves are numerical results, the dashed curves are the short-time analytics \eqref{xdotydot}.
}
\label{fig_gamma}
\end{figure}

The short-time result \eqref{xdotydot} indicates that a deflection in the $x$-direction is also possible without breaking chiral symmetry, if the initial wave packet has a phase shift $\gamma\neq \pi/4$. Such a phase shift between the electron and hole components could be induced by a voltage pulse. In Fig.\ \ref{fig_gamma} we show that the numerics confirms this analytical expectation.

\newpage

\begin{thebibliography}{99}
\bibitem{Kop91} N. B. Kopnin and M. M. Salomaa, \textit{Mutual friction in superfluid ${}^3$He: Effects of bound states in the vortex core},
Phys. Rev. B \textbf{44}, 9667 (1991) \doi{10.1103/PhysRevB.44.9667}.
\bibitem{Vol99} G. Volovik, \textit{Fermion zero modes on vortices in chiral superconductors}, JETP Lett. \textbf{70}, 609 (1999) \doi{10.1134/1.568223}.
\bibitem{Fu08} L. Fu and C. L. Kane, \textit{Superconducting proximity effect and Majorana fermions at the surface of a topological insulator}, Phys. Rev. Lett. \textbf{100}, 096407 (2008)  \doi{10.1103/PhysRevLett.100.096407}.
\bibitem{Frolov} S. Frolov, \url{https://espressospin.org/2012/04/18/zen-particle/}
\bibitem{Bee13} C. W. J. Beenakker, \textit{Search for Majorana fermions in superconductors}, Annu. Rev. Con. Mat. Phys. \textbf{4}, 113 (2013) \doi{10.1146/annurev-conmatphys-030212-184337}.
\bibitem{Das15} S. Das Sarma, M. Freedman, and C. Nayak  \textit{Majorana zero modes and topological quantum computation}, npj Quantum Inf. \textbf{1}, 15001 (2015) \doi{10.1038/npjqi.2015.1}.
\bibitem{Lut18} R. M. Lutchyn, E. P. A. M. Bakkers, L. P. Kouwenhoven, P. Krogstrup, C. M. Marcus, and Y. Oreg, \textit{Majorana zero modes in superconductor--semiconductor heterostructures}, Nature Rev. Mat. \textbf{3}, 52 (2018) \doi{10.1038/s41578-018-0003-1}.
\bibitem{Jac81} R. Jackiw and P. Rossi, \textit{Zero modes of the vortex-fermion system}, Nucl. Phys. B \textbf{190}, 681 (1981)  \doi{10.1016/0550-3213(81)90044-4}.
\bibitem{Noz66} P. Nozi\`{e}res and W. F. Vinen, \textit{The motion of flux lines in type II superconductors}, Phil. Mag. \textbf{14}, 667 (1966) \doi{10.1080/14786436608211964}.
\bibitem{Mak95} Yu. G. Makhlin and G. E. Volovik, \textit{Spectral flow in Josephson junctions and effective Magnus force}, JETP Lett. \textbf{62}, 941 (1995 ).
\bibitem{Sto96} M. L. Stone, \textit{Spectral flow, Magnus force, and mutual friction via the geometric optics limit of Andreev reflection}, Phys. Rev. B \textbf{54}, 13222 (1996) \doi{10.1103/PhysRevB.54.13222}.
\bibitem{Son97} E. B. Sonin \textit{Magnus force in superfluids and superconductors}, Phys. Rev. B \textbf{55}, 485 (1997) \doi{10.1103/PhysRevB.55.485}.
\bibitem{Pac21} M. J. Pacholski, G. Lemut, O. Ovdat, I. Adagideli, and C. W. J. Beenakker, \textit{Deconfinement of Majorana vortex modes produces a superconducting Landau level}, Phys. Rev. Lett. \textbf{126}, 226801 (2021) \doi{10.1103/PhysRevLett.126.226801}.
\bibitem{Zhu21} Zhen Zhu, Micha{\l} Papaj, Xiao-Ang Nie, Hao-Ke Xu, Yi-Sheng Gu, Xu Yang, Dandan Guan, Shiyong Wang, Yaoyi Li, Canhua Liu, Jianlin Luo, Zhu-An Xu, Hao Zheng, Liang Fu, and Jin-Feng Jia, \textit{Discovery of segmented Fermi surface induced by Cooper pair momentum}, Science \textbf{374},  1381 (2021) \doi{10.1126/science.abf107}. For a commentary on this experiment, see \doi{10.36471/JCCM_October_2020_02}.
\bibitem{Has10} M. Z. Hasan and C. L. Kane, \textit{Topological insulators}, Rev. Mod. Phys. \textbf{82}, 3045 (2010) \doi{10.1103/RevModPhys.82.3045}.
\bibitem{Qi11} X.-L. Qi and S.-C. Zhang, \textit{Topological insulators and superconductors}, Rev. Mod. Phys. \textbf{83}, 1057 (2011) \doi{10.1103/RevModPhys.83.1057}.
\bibitem{Tinkham} M. Tinkham, \textit{Introduction to Superconductivity} (Dover, New York, 2004).
\bibitem{Xia10} D. Xiao, M.-C. Chang, and Q. Niu, \textit{Berry phase effects on electronic properties}, Rev. Mod. Phys. \textbf{82}, 1959 (2010) \doi{10.1103/RevModPhys.82.1959}.
\bibitem{Wan21} Zhi Wang, Liang Dong, Cong Xiao, and Qian Niu, \textit{Berry curvature effects on quasiparticle dynamics in superconductors}, Phys. Rev. Lett. \textbf{126}, 187001 (2021) \doi{10.1103/PhysRevLett.126.187001}.
\bibitem{note1} The sign of the infinitesimal $y(0)$ for the equation of motion \eqref{dotydotky} is minus the sign of $K$ times the sign of the vorticity, in accord with the short-time dynamics \eqref{dotypi4}.
\bibitem{Cle75} J. R. Clem, \textit{Simple model for the vortex core in a type II superconductor}, J. Low Temp. Phys. \textbf{18}, 427 (1975) \doi{10.1007/BF00116134}.
\bibitem{Klo21} T. Kloss, J. Weston, B. Gaury, B. Rossignol, C. Groth, and X. Waintal, \textit{Tkwant: a software package for time-dependent quantum transport}, New J. Phys. \textbf{23}, 023025 (2021) \doi{10.1088/1367-2630/abddf7}.
\bibitem{zenodo} Our computer code will be made available in the Zenodo repository.
\bibitem{note2} For the comparison beween numerics and semiclassics in Fig.\ \ref{fig_colorscale} it makes no significant difference whether we take $P(y)\equiv K$ or include the near-field contribution from the circulating superfluid momentum, as in Eq.\ \eqref{deltapprofile}. 
\bibitem{Nay08} C. Nayak, S. H. Simon, A. Stern, M. Freedman, and S. Das Sarma, \textit{Non-Abelian anyons and topological quantum computation}, Rev. Mod. Phys. \textbf{80}, 1083 (2008) \doi{10.1103/RevModPhys.80.1083}.
\bibitem{Tu13} Hong-Hao Tu, Yi Zhang, and Xiao-Liang Qi, \textit{Momentum polarization: An entanglement measure of topological spin and chiral central charge}, Phys. Rev. B \textbf{88}, 195412 (2013) \doi{10.1103/PhysRevB.88.195412}.
\bibitem{Ari17} D. Ariad and E. Grosfeld, \textit{Signatures of the topological spin of Josephson vortices in topological superconductors}, Phys. Rev. B \textbf{95}, 161401(R) (2017) \doi{10.1103/PhysRevB.95.161401}.
\end{thebibliography}
\end{document}